\documentclass[conference]{IEEEtran}
\usepackage{amsmath}
\usepackage{booktabs}
\usepackage{graphicx}
\usepackage{listings}
\usepackage{xcolor}
\usepackage{soul}
\usepackage{url}
\usepackage{needspace}
\usepackage{tikz}
\usetikzlibrary{arrows.meta,positioning,fit,calc}
\usepackage{pgfplots}
\pgfplotsset{compat=1.18}

\newcommand{\lstheading}[1]{%
	\needspace{7\baselineskip}%
	\par\addvspace{0.6\baselineskip}%
	\refstepcounter{lstlisting}%
	{\normalfont\footnotesize \noindent
		Listing~\thelstlisting.\nobreakspace\nobreakspace #1\par}%
	\nopagebreak\vspace{3pt}%
}

\begin{document}
	\title{Silent Degradation during Auto-Vectorization of bf16/fp16 on RISC-V Under LLVM}
	
	\author{\IEEEauthorblockN{Vaishnavi Kouru}
		\IEEEauthorblockA{Department of Computer Science\\
			San Jose State University\\
			San Jose, CA, USA \\
			vkouru07@gmail.com}
		\and
		\IEEEauthorblockN{Robert Chun}
		\IEEEauthorblockA{Department of Computer Science\\
			San Jose State University\\
			San Jose, CA, USA \\
			robert.chun@sjsu.edu}}

	\maketitle 
	
	\begin{abstract}
		Low-precision numeric types such as 8-bit integers, bfloat16, and 
		half-precision floating point (fp16) are often used in AI inference 
		workloads to reduce memory bandwidth, and RISC-V's openly specified 
		Vector extension (RVV) is a natural target for compiling such 
		workloads with auto-vectorization. This paper examines whether 
		LLVM's RISC-V auto-vectorizer handles these narrow types equally well, 
		using a multiply-accumulate loop written for 
		fp32, int8, bf16, and fp16. Under a common baseline target 
		(\texttt{-march=rv64gcv}), int8 and fp32 vectorize cleanly, while bf16 
		and fp16 degrade to increasingly scalar-heavy fallback code, with fp16 
		falling back to per-element software conversion calls, unless 
		additional target-feature extensions (\texttt{zvfbfmin}/\texttt{zvfbfwma}, 
		\texttt{zvfhmin}) are enabled. Under QEMU user-mode 
		emulation, the baseline bf16 and fp16 kernels execute 17.0$\times$ 
		and 75.9$\times$ as many dynamic instructions as their 
		extension-enabled counterparts. Compiler diagnostics do report 
		this degradation, but do not distinguish, in their wording, between a 
		limitation that a compiler flag resolves entirely and one reflecting 
		an inherent conversion cost, leaving no direct signal toward 
		the fix. These results are supported by assembly-level instruction 
		analysis, by direct inspection of LLVM's optimization remarks, and by 
		dynamic instruction counts collected under emulation.
	\end{abstract}
	
	\section{Introduction}
	
	Using lower-precision numeric types, such as 8-bit integers and 16-bit 
	floating-point formats (e.g., bfloat16 and IEEE half-precision fp16), in place of 32-bit floating point is a growing trend as developers try to reduce 
	memory bandwidth and increase throughput in AI inference workloads 
	\cite{kalamkar2019,jacob2018}. 
	Model weights are commonly distributed in bf16 or fp16, and a developer 
	who keeps them in that format would reasonably expect the inner loops 
	that use them to run at least as fast as their fp32 versions, since 
	they move half as many bytes. Among these narrow types, integer and floating-point 
	formats are not vectorized equally well by default: as shown 
	below, an 8-bit integer type vectorizes as cleanly as standard 32-bit 
	floating point, while 16-bit floating-point types do not, unless 
	additional target-feature flags are enabled.
	
	RISC-V has also become a common choice for AI edge inference, partly because of its openly specified Vector extension (RVV) \cite{rvvspec}, which lets 
	compilers auto-vectorize loops without hand-written SIMD 
	intrinsics. Inference runtimes such as llama.cpp now include RVV 
	kernels \cite{ggml18199}, and recent work on RISC-V LLM inference 
	reports that the choice of compiler, and its support for ISA 
	extensions, changes throughput \cite{vseek2025}. 
	
	Developers targeting 
	RISC-V with a mainstream compiler may want to know whether auto-vectorization works
	equally well across the low-precision types AI workloads depend on, or 
	whether some types are handled worse than others without any clear sign 
	in the console output. 
	
	This paper examines that question with LLVM's auto-vectorizer, by compiling versions of a common multiply-accumulate loop (an operation of the 
	form \texttt{dst += a * b}, computed in one step) for the same RVV target, changing only the numeric type. 
	
	The compiler does not treat these types the same way: integer narrowing vectorizes 
	cleanly by default, while both floating-point narrow types degrade to 
	scalar-heavy fallback code unless additional target-feature flags 
	(\texttt{zvfbfmin}/\texttt{zvfbfwma} for bf16, \texttt{zvfhmin} for 
	fp16) are enabled, and the two floating-point types 
	degrade in different, increasingly severe ways. The degradation is 
	silent in a practical sense: the build succeeds, the output is 
	numerically correct, and no warning is printed unless optimization 
	remarks are requested. The compiler's  
	remarks, while accurate, do not distinguish between a 
	degradation that a compiler flag fixes and one that reflects 
	an inherent cost of the operation, so nothing in the text 
	points toward the fix.
	
	The extensions and their effects are documented, actively developed 
	LLVM and RISC-V features \cite{bf16spec,llvmpr115272,riscvsummit2025}. This study describes, 
	at the assembly and diagnostic level, how 
	vectorization quality degrades across three AI-relevant narrow types 
	under the same method, measures the resulting difference in 
	executed instructions under emulation, traces the degradation to the 
	stage of the LLVM pipeline where it is decided, and identifies a 
	specific gap in how compiler remarks report that degradation to 
	developers.
	
	\section{Background}
	
	\subsection{Narrow types in AI inference}
	
	bf16 keeps fp32's 8-bit exponent and truncates the mantissa to 7 bits, 
	so an fp32 checkpoint can be converted to bf16 without rescaling; a 
	bf16 value is exactly the upper 16 bits of the corresponding fp32 bit 
	pattern \cite{kalamkar2019}. fp16 uses a 5-bit exponent and a 10-bit 
	mantissa, giving more precision over a narrower range. 
	int8 is used for quantized inference, where fp32 weights are scaled 
	into a small integer range \cite{jacob2018}.
	
	In all three cases, inference kernels usually keep the narrow type only 
	for storage and accumulate in a wider type: 16-bit floating-point 
	inputs are widened to fp32 before the multiply-accumulate 
	\cite{kalamkar2019}, and int8 products are accumulated in int32 
	\cite{jacob2018}.
	
	\subsection{Why the multiply-accumulate loop matters}
	\label{sec:mac}
	
	Most of the arithmetic in a neural network comes from a few layer 
	types. A fully connected layer computes each output as a weighted sum 
	of its inputs, $y_j = \sum_i W_{ji} x_i$. A convolution computes the 
	same kind of sum over a small window of its input, and an attention 
	layer computes it when comparing queries with keys and when weighting 
	values. All of these are carried out as matrix multiplication, and 
	matrix multiplication is made of one repeated step: multiply two 
	numbers and add the product to a running total. This step is the 
	multiply-accumulate.
	
	Multiplying an $M \times K$ matrix by a $K \times N$ matrix performs 
	this step $M K N$ times, so the inner loop that performs it accounts 
	for nearly all of the kernel's running time. Any extra work inside 
	that loop, such as converting each operand from bf16 to fp32 one 
	element at a time, is repeated just as many times. The classifier 
	layer of MobileNetV2, whose weights are used in 
	Section~\ref{sec:exec-method}, maps 1280 inputs to 1001 outputs and 
	so performs about 1.28 million multiply-accumulates for each image.
	
	The loop studied here is this step in its simplest form: it loads two 
	narrow operands, widens them to the accumulator type, multiplies them, 
	and adds the result into a wider destination. Optimized matrix 
	multiplication kernels arrange their loops differently, splitting the 
	work into tiles that fit in registers and cache, but the innermost 
	loop of each tile performs the same load, widen, multiply, and add 
	sequence over a vector of elements. Whether the compiler can vectorize 
	this sequence for a given type therefore decides whether the larger 
	kernel can be vectorized well for that type. The loop is also small 
	enough that its full assembly can be read and counted by hand, which 
	is not practical for a complete matrix multiplication kernel.
	
	These loops are 
	usually not written by the person training or deploying the model; 
	they are part of inference runtimes and math libraries compiled from C or 
	C++ source for the target device \cite{ggml18199}, and the \texttt{-march} string used 
	for that build decides which of the code paths shown in this paper is 
	taken.
	
	\subsection{The RISC-V Vector extension (RVV)}
	
	RVV is RISC-V's standard extension for vector (SIMD-style) computation \cite{rvvspec}. 
	Unlike fixed-width SIMD extensions such as x86's AVX, where the vector 
	register width is fixed and known at compile time, RVV is 
	\emph{scalable}: the actual number of elements processed per 
	instruction is determined by the hardware at runtime via the 
	\texttt{vsetvli} instruction, which the compiler emits at the start of 
	a vectorized loop to query and configure the active vector length. 
	
	Two parameters control how a given operation is vectorized: 
	\textbf{SEW} (Selected Element Width), the bit-width of each individual 
	element (8, 16, 32, or 64 bits), and \textbf{LMUL} (vector register 
	group multiplier), which controls how many physical vector registers 
	are grouped together to act as one larger logical vector register. A 
	higher LMUL allows more data to be processed per instruction at the 
	cost of using up more of the register file, reducing how many 
	independent vector values can be live at once. The 
	compiler's cost model must estimate the tradeoff correctly to make good 
	vectorization decisions, particularly for operations such as widening 
	multiply-accumulate that produce a wider result type than their inputs.
	
	\subsection{Target-feature extensions for narrow types}
	
	RVV's base vector extension (\texttt{v}) guarantees vector support for 
	standard integer and floating-point types, and integer narrowing 
	operations (e.g. int8 inputs widening to an int32 accumulator) are 
	covered by this base support. Narrow floating-point formats are 
	handled differently: because not all RISC-V implementations support 
	native bf16 or fp16 arithmetic, RVV specifies them as separate, 
	optional extensions layered on top of \texttt{v}: \texttt{Zvfbfmin} 
	(minimal bf16 vector load/convert support) and \texttt{Zvfbfwma} 
	(bf16 widening multiply-add) for bfloat16 \cite{bf16spec}, and 
	\texttt{Zvfhmin} for fp16. A compiler targeting only the base \texttt{v} extension cannot 
	assume these are available, and must fall back to other ways of 
	generating code for loops that use these types. The RVA23U64 
	application profile makes \texttt{Zvfhmin} mandatory but leaves 
	\texttt{Zvfbfmin} and \texttt{Zvfbfwma} optional \cite{rva23}.
	
	\subsection{Compiler auto-vectorization and diagnostics}
	\label{sec:diagnostics}
	
	LLVM's Loop Vectorizer analyzes each loop and, using a target-specific 
	cost model, decides whether vectorizing is likely to be profitable, 
	and at what width. Clang reports these decisions through 
	optimization remarks: \texttt{-Rpass=loop-vectorize} reports 
	successful vectorization and its width, \texttt{-Rpass-missed} reports 
	loops that failed to vectorize, and \texttt{-Rpass-analysis} reports 
	the cost-model reasoning behind these decisions. These remarks are the 
	main way a developer, without reading the generated 
	assembly, can learn how the compiler handled a given loop.
	
	\subsection{Where the decision is made}
	
	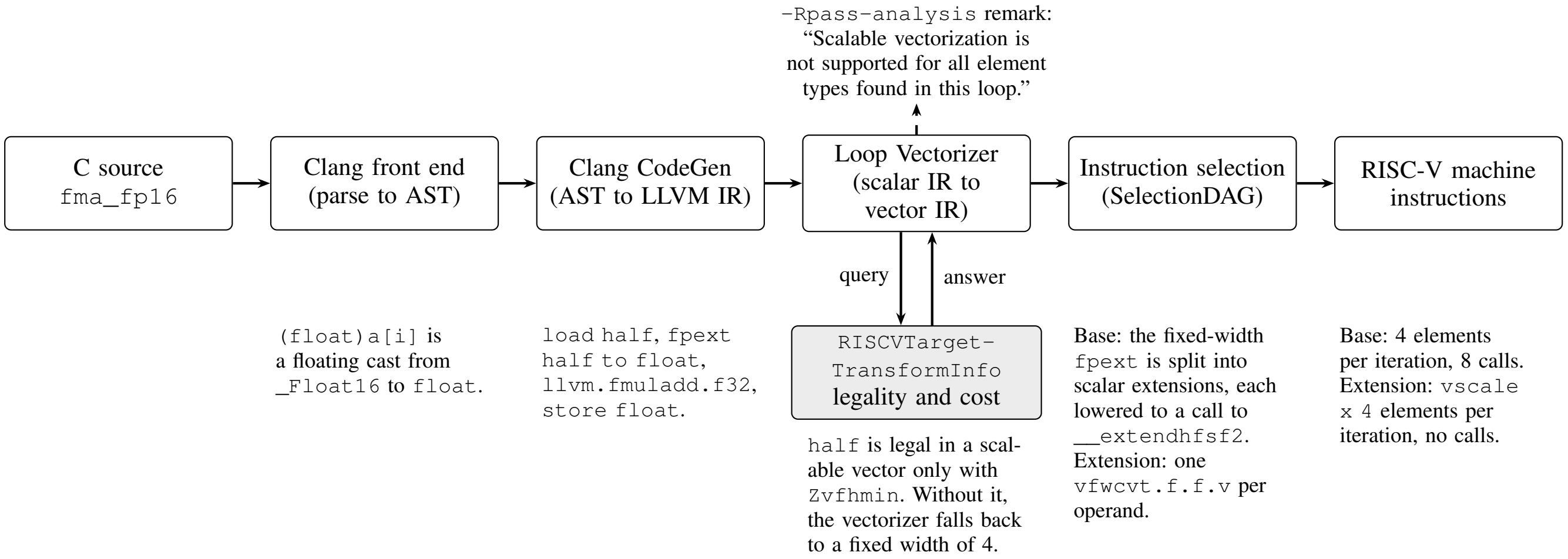
\begin{figure*}[t]
		\centering
		\begin{tikzpicture}[
	font=\footnotesize,
	stage/.style={draw, rounded corners=2pt, align=center, text width=2.35cm,
		minimum height=1.05cm, inner sep=3pt},
	tti/.style={stage, fill=gray!15, text width=2.6cm},
	note/.style={align=left, text width=2.45cm, font=\scriptsize, inner sep=1pt},
	arr/.style={-{Stealth[length=4pt]}, thick},
	node distance=0.42cm
	]
	\node[stage] (src) {C source\\\texttt{fma\_fp16}};
	\node[stage, right=of src] (ast) {Clang front end\\(parse to AST)};
	\node[stage, right=of ast] (ir) {Clang CodeGen\\(AST to LLVM IR)};
	\node[stage, right=of ir] (lv) {Loop Vectorizer\\(scalar IR to vector IR)};
	\node[stage, right=of lv] (isel) {Instruction selection\\(SelectionDAG)};
	\node[stage, right=of isel] (mi) {RISC-V machine instructions};
	
	\foreach \a/\b in {src/ast, ast/ir, ir/lv, lv/isel, isel/mi}
	\draw[arr] (\a) -- (\b);
	
	\node[tti, below=1.05cm of lv] (tti) {{\scriptsize\texttt{RISCVTarget-}\\\texttt{TransformInfo}}\\legality and cost};
	\draw[arr] ([xshift=-5pt]lv.south) -- node[left, font=\scriptsize, align=right] {query} ([xshift=-5pt]tti.north);
	\draw[arr] ([xshift=5pt]tti.north) -- node[right, font=\scriptsize, align=left] {answer} ([xshift=5pt]lv.south);
	
	\node[note, above=0.35cm of lv, text width=3.3cm, align=center] (rm)
	{\texttt{-Rpass-analysis} remark: ``Scalable vectorization is not supported for all element types found in this loop.''};
	\draw[arr, dashed] (lv.north) -- (rm.south);
	
	\node[note, below=1.05cm of ast, anchor=north] {\texttt{(float)a[i]} is a floating cast from \texttt{\_Float16} to \texttt{float}.};
	\node[note, below=1.05cm of ir, anchor=north] {\texttt{load half}, \texttt{fpext half to float}, \texttt{llvm.fmuladd.f32}, \texttt{store float}.};
	\node[note, below=0.15cm of tti, anchor=north] {\texttt{half} is legal in a scalable vector only with \texttt{Zvfhmin}. Without it, the vectorizer falls back to a fixed width of 4.};
	\node[note, below=1.05cm of isel, anchor=north] {Base: the fixed-width \texttt{fpext} is split into scalar extensions, each lowered to a call to \texttt{\_\_extendhfsf2}. Extension: one \texttt{vfwcvt.f.f.v} per operand.};
	\node[note, below=1.05cm of mi, anchor=north] {Base: 4 elements per iteration, 8 calls. Extension: \texttt{vscale x 4} elements per iteration, no calls.};
\end{tikzpicture}
		\caption{Path of \texttt{fma\_fp16} through Clang and LLVM. The Loop Vectorizer queries \texttt{RISCVTargetTransformInfo} for whether each element type is legal in a scalable vector and for the cost of each operation. Without \texttt{Zvfhmin}, \texttt{half} is rejected, the vectorizer falls back to a fixed width, and instruction selection later expands each conversion into a library call. bf16 follows the same path, with \texttt{Zvfbfmin} in place of \texttt{Zvfhmin} and integer shifts in place of calls.}
		\label{fig:llvm-pipeline}
	\end{figure*}
	
	Fig.~\ref{fig:llvm-pipeline} shows the stages a kernel passes through. 
	Clang parses the source into an AST, in which each \texttt{(float)} 
	cast is a floating-point conversion node, and lowers it to LLVM IR as 
	a narrow load, an \texttt{fpext} to \texttt{float}, and an 
	\texttt{llvm.fmuladd.f32} call. The Loop Vectorizer then chooses a 
	vector width. Before choosing a scalable width, it asks the target, 
	through \texttt{RISCVTargetTransformInfo}, whether every element type 
	in the loop is legal in a scalable vector. On RISC-V, \texttt{half} 
	is legal only when \texttt{Zvfhmin} is enabled and \texttt{bfloat} 
	only when \texttt{Zvfbfmin} is enabled. If any type is rejected, 
	scalable vectorization is ruled out, the remark ``Scalable 
	vectorization is not supported for all element types found in this 
	loop'' is emitted, and the vectorizer falls back to a fixed width 
	(here 4). Treating \texttt{half} and \texttt{bfloat} as legal under 
	\texttt{Zvfhmin} and \texttt{Zvfbfmin} alone, rather than only under 
	the full \texttt{Zvfh} extension, was added to LLVM in November 2024 
	\cite{llvmpr115272}.
	
	The fixed-width vector IR still contains a vector \texttt{fpext} from 
	16-bit to 32-bit elements. Instruction selection has no single RISC-V 
	instruction for this conversion without the extension, so it splits 
	the operation into scalar conversions. For bf16, each scalar 
	conversion is an integer shift that places the 16 bits in the upper 
	half of an fp32 register, followed by reinsertion into a vector. For 
	fp16, the base target also lacks scalar half-precision support, so 
	each scalar conversion becomes a call to the runtime library routine 
	\texttt{\_\_extendhfsf2}. With the extension enabled, the same IR 
	lowers to \texttt{vfwcvt.f.f.v} for fp16, and the widening and 
	multiply-accumulate combine into \texttt{vfwmaccbf16.vv} for bf16.
	
	\section{Related Work}
	
	Prior work has studied LLVM's RISC-V auto-vectorization support 
	using general-purpose benchmark suites. A 2025 study by researchers at 
	LLNL evaluated vectorization coverage across the Test Suite for 
	Vectorizing Compilers (TSVC), comparing LLVM and GCC across RVV and 
	ARM SVE targets \cite{llnl2025}. A separate 2025 study evaluated 
	auto-vectorization performance on physical RVV-capable RISC-V hardware, 
	using TSVC loops alongside several real-world applications, and found 
	cases where LLVM failed to vectorize loops that GCC handled 
	successfully \cite{pdp2025}. More recently, a 2026 study compared 
	GCC~15 and Clang~21 code generation across scientific and ML 
	workloads, including SGEMM/DGEMM, and found that default LMUL 
	selection is close to optimal in most cases, identifying predication 
	overhead and strided memory access, rather than LMUL choice, as the 
	primary sources of remaining performance gaps \cite{closerinthegap2026}.
	
	These studies evaluate general-purpose or full-precision numeric 
	workloads and do not specifically isolate the behavior of low-precision, 
	AI-relevant types such as bf16, fp16, or int8. Separately, LLVM's own 
	RISC-V maintainers have described ongoing work to support vectorized 
	code generation for the \texttt{Zvfbfmin}/\texttt{Zvfbfwma} and 
	\texttt{Zvfhmin} extensions, noting that widened bf16/fp16 operations 
	need extra care in the cost model because of their effect on LMUL and 
	register pressure \cite{riscvsummit2025}. This shows that the 
	mechanism examined in this study is a known, active 
	area of compiler development, but its effect on generated 
	code quality, and on the clarity of the resulting compiler diagnostics, 
	has not been measured in the studies above. The change that made 
	\texttt{half} and \texttt{bfloat} legal for scalable vectorization 
	under the minimal extensions was tested by compiling TSVC with bf16 
	element types \cite{llvmpr115272}, but that work concerns correctness 
	of the extension-enabled path rather than the cost of the baseline one.
	
	On the inference side, a 2025 study porting llama.cpp to the many-core SG2042 RISC-V 
	CPU found that Clang~19 outperformed GCC~13.2 by 34\% in token 
	generation, and attributed the gap in part to ISA extension support 
	\cite{vseek2025}. llama.cpp's bf16 RVV kernels using 
	\texttt{Zvfbfwma} have so far been tested only under QEMU 
	\cite{ggml18199}. These reports show that toolchain configuration 
	affects inference performance on RISC-V, but do not isolate the 
	per-type effect measured here.
	
	Research on compiler diagnostics has found that developers do read 
	compiler error messages, and that messages which are harder to read 
	lead to worse task performance \cite{barik2017}. That work 
	concerns error messages; this study applies the same concern to 
	optimization remarks.
	
	This study differs from prior work in two ways: it focuses on 
	narrow, AI-relevant types rather than general-purpose 
	benchmark suites, and it evaluates not only whether vectorization 
	succeeds but how clearly the compiler explains the result 
	when it does not succeed cleanly.
	
	\section{Methodology}
	
	\begin{figure}[t]
		\centering
		\includegraphics[width=\columnwidth]{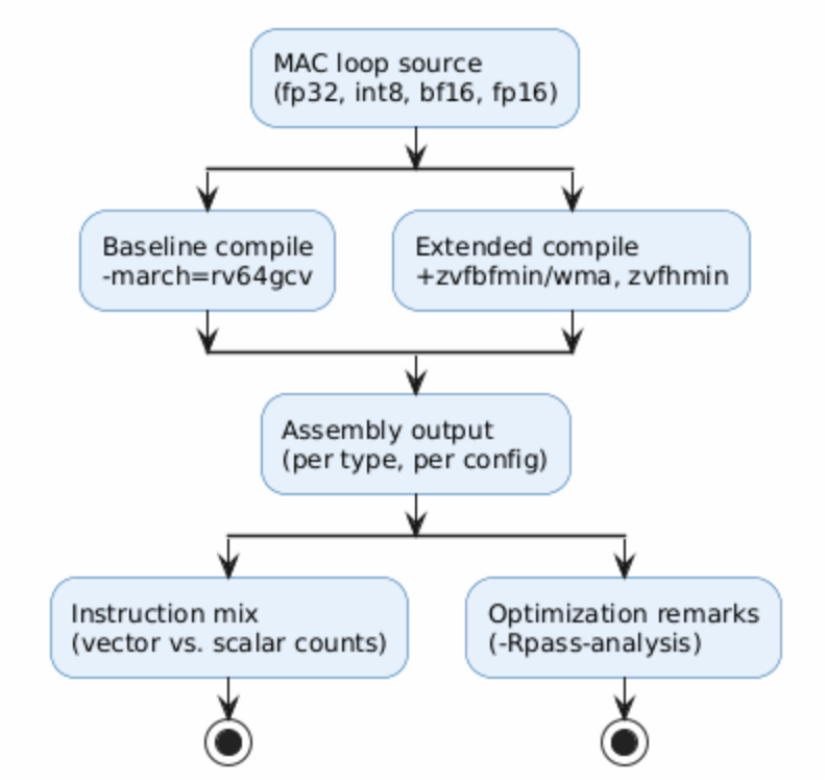}
		\caption{Methodology: source variants, baseline vs.\ extended target-feature compilation, and the two static analysis channels. Execution measurement is described in Section~\ref{sec:exec-method}.}
		\label{fig:pipeline}
	\end{figure}
	
	All kernels were compiled with Clang~22 (22.1.8) for 
	\texttt{riscv64-unknown-linux-gnu}, the same version used for 
	the measured binaries in Section~\ref{sec:exec-method}. The analysis 
	was also repeated with Clang~23.1.3, the current release at the time 
	of writing, and with an LLVM~24 development build from August 2026. 
	Both produce the same remarks and the same fallback code; under 
	Clang~23.1.3, every vector loop body reported below has the same 
	instruction count except the baseline fp16 loop, which is one 
	instruction shorter. Common flags across all runs were 
	\texttt{-O3 -mabi=lp64d}, plus diagnostic flags 
	\texttt{-Rpass=loop-vectorize -Rpass-missed=loop-vectorize 
		-Rpass-analysis=loop-vectorize} to show the vectorizer's cost-model 
	reasoning. Each kernel was compiled under a baseline target 
	(\texttt{-march=rv64gcv}) and, where applicable, again with a 
	narrow-type extension added, to isolate that extension's effect on 
	generated code.
	
	This study evaluates a single loop shape, an elementwise fused 
	multiply-accumulate over an array, instantiated across four input 
	types, along with one reduction-shaped kernel (\texttt{dot\_int8}, discussed 
	below). The elementwise shape was chosen because it matches 
	the innermost loop of common AI operations, as described in 
	Section~\ref{sec:mac}, while staying small enough to 
	read the full assembly by hand.
	
	\subsection{Base loop}
	
	\lstheading{Base FMA loop template, instantiated per type below}
	\begin{lstlisting}[language=C]
		void fma_<TYPE>(<ACC_TYPE> *dst, const <TYPE> *a, const <TYPE> *b, int n) {
			for (int i = 0; i < n; i++) {
				dst[i] += (<ACC_TYPE>)a[i] * (<ACC_TYPE>)b[i];
			}
		}
	\end{lstlisting}
	
	\subsection{fp32 (\texttt{TYPE = float}, \texttt{ACC\_TYPE = float}), reference case}
	
	\lstheading{fp32 instantiation, used as a reference for standard-width vectorization}
	\begin{lstlisting}[language=C]
		void fma_f32(float *dst, const float *a, const float *b, int n) {
			for (int i = 0; i < n; i++) {
				dst[i] += a[i] * b[i];
			}
		}
	\end{lstlisting}
	Flags: \texttt{-march=rv64gcv} (no additional extension evaluated, since 
	this type is fully supported by the base vector extension and serves 
	as a reference point for standard-width vectorization quality).
	
	\subsection{int8 (\texttt{TYPE = signed char}, \texttt{ACC\_TYPE = int})}
	
	Two int8 kernels were evaluated. The first, \texttt{dot\_int8}, is 
	structured as a reduction (accumulating into a scalar sum) rather than 
	an elementwise array update, since integer dot products are the 
	usual form for int8 quantized inference (e.g., in 
	quantized GEMM/dot-product microkernels):
	
	\lstheading{int8 instantiation, as a dot product (reduction)}
	\begin{lstlisting}[language=C]
		int dot_int8(const signed char *a, const signed char *b, int n) {
			int sum = 0;
			for (int i = 0; i < n; i++) {
				sum += (int)a[i] * (int)b[i];
			}
			return sum;
		}
	\end{lstlisting}
	Flags: \texttt{-march=rv64gcv} (no additional extension evaluated; 
	see Results \S\ref{sec:results}).
	
	Because \texttt{dot\_int8}'s reduction structure differs 
	from the elementwise \texttt{fma\_*} kernels used for the other three 
	types, a second, elementwise-only int8 kernel, \texttt{fma\_int8}, was 
	also evaluated, matching the base loop template exactly:
	
	\lstheading{int8 instantiation, elementwise (topology-matched to fma\_f32/bf16/fp16)}
	\begin{lstlisting}[language=C]
		void fma_int8(int *dst, const signed char *a, const signed char *b, int n) {
			for (int i = 0; i < n; i++) {
				dst[i] += (int)a[i] * (int)b[i];
			}
		}
	\end{lstlisting}
	Flags: \texttt{-march=rv64gcv}.
	
	\texttt{dot\_int8} is kept because it reflects the usual form of 
	this operation in practice, while \texttt{fma\_int8} gives a direct, 
	structure-matched comparison with the other three \texttt{fma\_*} 
	kernels.
	
	\needspace{12\baselineskip}
	\subsection{bf16 (\texttt{TYPE = \_\_bf16}, \texttt{ACC\_TYPE = float})}
	
	\lstheading{bf16 instantiation}
	\begin{lstlisting}[language=C]
		void fma_bf16(float *dst, const __bf16 *a, const __bf16 *b, int n) {
			for (int i = 0; i < n; i++) {
				dst[i] += (float)a[i] * (float)b[i];
			}
		}
	\end{lstlisting}
	Flags: \texttt{-march=rv64gcv} (baseline) and 
	\texttt{-march=rv64gcv\_zvfbfmin\_zvfbfwma} (extension added).
	
	\needspace{12\baselineskip}
	\subsection{fp16 (\texttt{TYPE = \_Float16}, \texttt{ACC\_TYPE = float})}
	
	\lstheading{fp16 instantiation}
	\begin{lstlisting}[language=C]
		void fma_fp16(float *dst, const _Float16 *a, const _Float16 *b, int n) {
			for (int i = 0; i < n; i++) {
				dst[i] += (float)a[i] * (float)b[i];
			}
		}
	\end{lstlisting}
	Flags: \texttt{-march=rv64gcv} (baseline) and 
	\texttt{-march=rv64gcv\_zvfhmin} (extension added).
	
	\subsection{Summary of runs}
	
	\begin{table}[h]
		\centering
		\caption{All compilation runs performed}
		\label{tab:runs}
		\begin{tabular}{p{1.4cm} p{3.9cm} p{2.3cm}} 
			\toprule
			\textbf{Kernel} & \textbf{Flags} & \textbf{Role} \\
			\midrule
			\texttt{dot\_int8} & \texttt{rv64gcv} & Baseline (reduction) \\
			\texttt{fma\_int8} & \texttt{rv64gcv} & Baseline (elementwise) \\
			\texttt{fma\_f32} & \texttt{rv64gcv} & Baseline (reference) \\
			\texttt{fma\_bf16} & \texttt{rv64gcv} & Baseline \\
			\texttt{fma\_bf16} & \texttt{rv64gcv\_zvfbfmin\_zvfbfwma} & Extension added \\
			\texttt{fma\_fp16} & \texttt{rv64gcv} & Baseline \\
			\texttt{fma\_fp16} & \texttt{rv64gcv\_zvfhmin} & Extension added \\
			\bottomrule
		\end{tabular}
	\end{table}
	
	\subsection{Execution measurement}
	\label{sec:exec-method}
	
	To measure the effect of the generated code rather than infer it from 
	static listings, each of the seven configurations in 
	Table~\ref{tab:runs} was compiled into a standalone benchmark and run 
	under emulation. Binaries were built on an Ubuntu Linux host with Clang~22.1.8 
	for \texttt{riscv64-unknown-linux-gnu}, using the same \texttt{-O3 
	-mabi=lp64d} and \texttt{-march} strings as above, statically linked 
	against the Ubuntu RISC-V cross glibc. They were executed with QEMU~9.1.0 
	\cite{bellard2005} in user mode (\texttt{qemu-riscv64}), configured as an \texttt{rv64} 
	CPU with the vector extension, VLEN = 256, and ELEN = 64, with 
	\texttt{zvfbfmin} and \texttt{zvfbfwma}, or \texttt{zvfhmin}, 
	also enabled for the extension builds. Emulation is also how 
	llama.cpp's bf16 RVV kernels were tested, since their developers did 
	not have hardware implementing \texttt{Zvfbfwma} \cite{ggml18199}.
	
	Every kernel operates on arrays of $N = 147{,}456$ elements 
	($384 \times 384$). Input values were taken from the largest weight 
	tensor (the classifier weight matrix) of a pretrained MobileNetV2 
	\cite{sandler2018} checkpoint and converted once from fp32 to each type: bf16 and fp16 by 
	round-to-nearest conversion, and int8 by symmetric scaling to 
	$[-127, 127]$. All kernels therefore consume the same underlying 
	values, and the second operand is a copy of the first. Each binary calls 
	its kernel 100 times over the full array (14.7 million elements in 
	total) and was run 5 times; medians are reported.
	
	Two quantities were recorded. The primary one is the dynamic guest 
	instruction count, collected with QEMU's \texttt{libinsn} plugin, 
	which counts every RISC-V instruction the program executes, including 
	startup, file input, and initialization, which are common to all 
	binaries and small relative to the kernel. The secondary one is host 
	wall-clock time of the timed kernel region, measured with 
	\texttt{clock\_gettime} inside the guest. Instruction count is used for 
	comparison because QEMU translates each guest instruction into host code 
	or a software helper routine whose host cost varies by instruction and 
	does not correspond to the cost of that instruction on any RISC-V core.
	
	The RISC-V runtime library on the build host does not provide 
	\texttt{\_\_extendhfsf2}, which the baseline fp16 binary calls. A 
	portable C implementation of the routine (bit manipulation on the 
	16-bit input, with handling for zero, subnormal, infinite, and NaN 
	inputs) was linked into both fp16 binaries; only the baseline binary 
	calls it.
	
	\section{Results}
	\label{sec:results}
	
	\subsection{Determining vectorization outcomes}
	
	Vectorization outcomes are judged using two signals. 
	First, Clang's \texttt{-Rpass=loop-vectorize} remark directly reports 
	whether a loop was vectorized and at what width. A fixed integer 
	(e.g. \texttt{4}) indicates compile-time-fixed vectorization, while 
	\texttt{vscale x N} indicates scalable, vector-length-agnostic 
	vectorization. This remark alone, however, does not capture 
	\emph{quality}: a loop may be classified as vectorized while still 
	containing substantial scalar overhead. For this reason, the generated 
	assembly is also inspected for the share of the loop body 
	taken up by real vector instructions (\texttt{vle*.v}, 
	\texttt{vfmacc.vv}, \texttt{vwmacc.vv}, etc.) versus scalar 
	reconstruction work (per-element scalar loads, \texttt{vslideup.vi} 
	sequences used to manually assemble vector operands, or calls to 
	software conversion routines such as \texttt{\_\_extendhfsf2}). This 
	distinction is what allows \texttt{fma\_bf16} and 
	\texttt{fma\_fp16} under baseline flags to be characterized as 
	\emph{degraded} rather than simply vectorized or not vectorized, a 
	classification the remark text alone does not support.
	
	\subsection{Baseline vectorization outcomes}
	
	Table~\ref{tab:baseline} summarizes vectorization outcomes for all 
	kernels under the baseline target \texttt{-march=rv64gcv}.
	
	\begin{table*}[t]
		\centering
		\caption{Vectorization outcome under baseline \texttt{-march=rv64gcv}}
		\label{tab:baseline}
		\begin{tabular}{lccl}
			\toprule
			\textbf{Kernel} & \textbf{Vectorized?} & \textbf{Width} & \textbf{Notes} \\
			\midrule
			\texttt{dot\_int8} & Yes (clean) & \texttt{vscale x 4} & No degradation remarks \\
			\texttt{fma\_int8} & Yes (clean) & \texttt{vscale x 4} & No degradation remarks \\
			\texttt{fma\_f32} & Yes (clean) & \texttt{vscale x 4} & No degradation remarks \\
			\texttt{fma\_bf16} & Yes (degraded) & 4 (fixed) & Manual vector reconstruction \\
			\texttt{fma\_fp16} & Yes (degraded) & 4 (fixed) & Scalar \texttt{\_\_extendhfsf2} calls \\
			\bottomrule
		\end{tabular}
	\end{table*}
	
	\subsection{Effect of narrow-type extensions}
	
	Table~\ref{tab:extension-fix} shows the effect of adding the relevant 
	narrow-type extension for the two kernels that degraded under the 
	baseline.
	
	\begin{table*}[t]
		\centering
		\caption{Effect of adding minimal narrow-type vector extensions}
		\label{tab:extension-fix}
		\begin{tabular}{lccc}
			\toprule
			\textbf{Kernel} & \textbf{Extension added} & \textbf{Width (before $\to$ after)} & \textbf{Degradation removed?} \\
			\midrule
			\texttt{fma\_bf16} & \texttt{zvfbfmin+zvfbfwma} & 4 $\to$ \texttt{vscale x 4} & Yes \\
			\texttt{fma\_fp16} & \texttt{zvfhmin} & 4 $\to$ \texttt{vscale x 4} & Yes \\
			\bottomrule
		\end{tabular}
	\end{table*}
	
	For \texttt{fma\_fp16}, the compiler's diagnostics show 
	this distinction directly. Under the baseline flags, the vectorizer 
	reports:
	
	\begin{quote}
		\texttt{remark: Scalable vectorization is not supported for all 
			element types found in this loop.}
	\end{quote}
	
	alongside:
	
	\begin{quote}
		\texttt{remark: floating point conversion changes vector width. Mixed 
			floating point precision requires an up/down cast that will negatively 
			impact performance.}
	\end{quote}
	
	After adding \texttt{zvfhmin}, the first remark disappears entirely 
	and vectorization width changes from a fixed \texttt{4} to 
	\texttt{vscale x 4}, confirming that this remark corresponded to a 
	limitation a flag can fix. The second remark, however, appears 
	unchanged in both runs, since it reflects an inherent cost of the 
	fp16-to-fp32 conversion rather than a missing hardware capability. 
	Both remarks are emitted through the same 
	\texttt{-Rpass-analysis=loop-vectorize} channel with similar 
	phrasing, so nothing in the text tells the developer that one can be 
	fixed with a compiler flag and the other cannot.
	
	\subsection{Instruction-level comparison}
	
	\begin{figure*}[t]
		\centering
		\begin{tikzpicture}[
	font=\scriptsize,
	box/.style={draw, align=center, text width=4.1cm, minimum height=0.42cm, inner sep=2pt},
	prod/.style={box},
	over/.style={box, fill=gray!30},
	ctrl/.style={box, dashed},
	hdr/.style={align=center, text width=5.0cm, font=\footnotesize},
	foot/.style={align=center, text width=5.0cm, font=\footnotesize},
	node distance=2pt
	]
	
	\node[hdr] (ha) at (0,0) {\textbf{(a) Native vector path}\\fp16 with \texttt{zvfhmin}};
	\node[ctrl, below=6pt of ha] (a1) {\texttt{vsetvli}: vl = min(remaining, VLMAX)};
	\node[prod, below=of a1] (a2) {\texttt{vle16.v} $\times$2 (load $a$, $b$)};
	\node[prod, below=of a2] (a3) {\texttt{vfwcvt.f.f.v} $\times$2 (widen to fp32)};
	\node[prod, below=of a3] (a4) {\texttt{vle32.v} (load dst)};
	\node[prod, below=of a4] (a5) {\texttt{vfmacc.vv}};
	\node[prod, below=of a5] (a6) {\texttt{vse32.v} (store dst)};
	\node[ctrl, below=of a6] (a7) {pointer and counter updates, branch};
	
	\node[hdr] (hb) at (5.3,0) {\textbf{(b) Scalar reconstruction path}\\bf16, baseline};
	\node[over, below=6pt of hb] (b1) {\texttt{lhu} $\times$8 (scalar 16-bit loads)};
	\node[over, below=of b1] (b2) {shift $\times$8 (bf16 bits to fp32 position)};
	\node[over, below=of b2] (b3) {\texttt{vmv.s.x} $\times$8 (scalar to vector)};
	\node[over, below=of b3] (b4) {\texttt{vslideup.vi} $\times$6 (assemble 4 lanes)};
	\node[prod, below=of b4] (b5) {\texttt{vle32.v} (load dst)};
	\node[prod, below=of b5] (b6) {\texttt{vfmacc.vv}};
	\node[prod, below=of b6] (b7) {\texttt{vse32.v} (store dst)};
	\node[ctrl, below=of b7] (b8) {pointer and counter updates, branch};
	
	\node[hdr] (hc) at (10.6,0) {\textbf{(c) Software library path}\\fp16, baseline};
	\node[over, below=6pt of hc] (c1) {spill 14 registers (once per function call)};
	\node[font=\scriptsize, below=5pt of c1] (glab) {repeated $\times$8 per iteration};
	\node[over, below=1pt of glab] (c2) {move operand to argument register};
	\node[over, below=of c2] (c3) {\texttt{call \_\_extendhfsf2} (software conversion)};
	\node[over, below=of c3] (c4) {\texttt{vs1r.v} / \texttt{vl1r.v} (vector spill, reload)};
	\node[draw, dotted, thick, fit=(glab)(c2)(c4), inner sep=2pt] (grp) {};
	\node[over, below=5pt of c4] (c5) {move results into vector registers};
	\node[prod, below=of c5] (c6) {\texttt{vle32.v}, \texttt{vfmacc.vv}, \texttt{vse32.v}};
	\node[ctrl, below=of c6] (c7) {pointer and counter updates, branch};
	
	\path (c7.south) ++(0,-0.25) coordinate (fy);
	\node[foot, anchor=north] at (ha.center |- fy) {vl elements per iteration (set at run time)\\Measured: 0.82 instructions/element};
	\node[foot, anchor=north] at (hb.center |- fy) {4 elements per iteration (fixed)\\Measured: 9.76 instructions/element};
	\node[foot, anchor=north] at (hc.center |- fy) {4 elements per iteration (fixed)\\Measured: 62.5 instructions/element};
\end{tikzpicture}
		\caption{Per-iteration loop bodies for the three code paths observed. White boxes are productive vector work, gray boxes are overhead present only because the extension is absent, and dashed boxes are loop control. Path (a) is shown for fp16; the bf16 extension build is similar, with a single \texttt{vfwmaccbf16.vv} replacing the two conversions and the \texttt{vfmacc.vv}. Measured values are dynamic instructions per element from Table~\ref{tab:exec}.}
		\label{fig:exec-paths}
	\end{figure*}
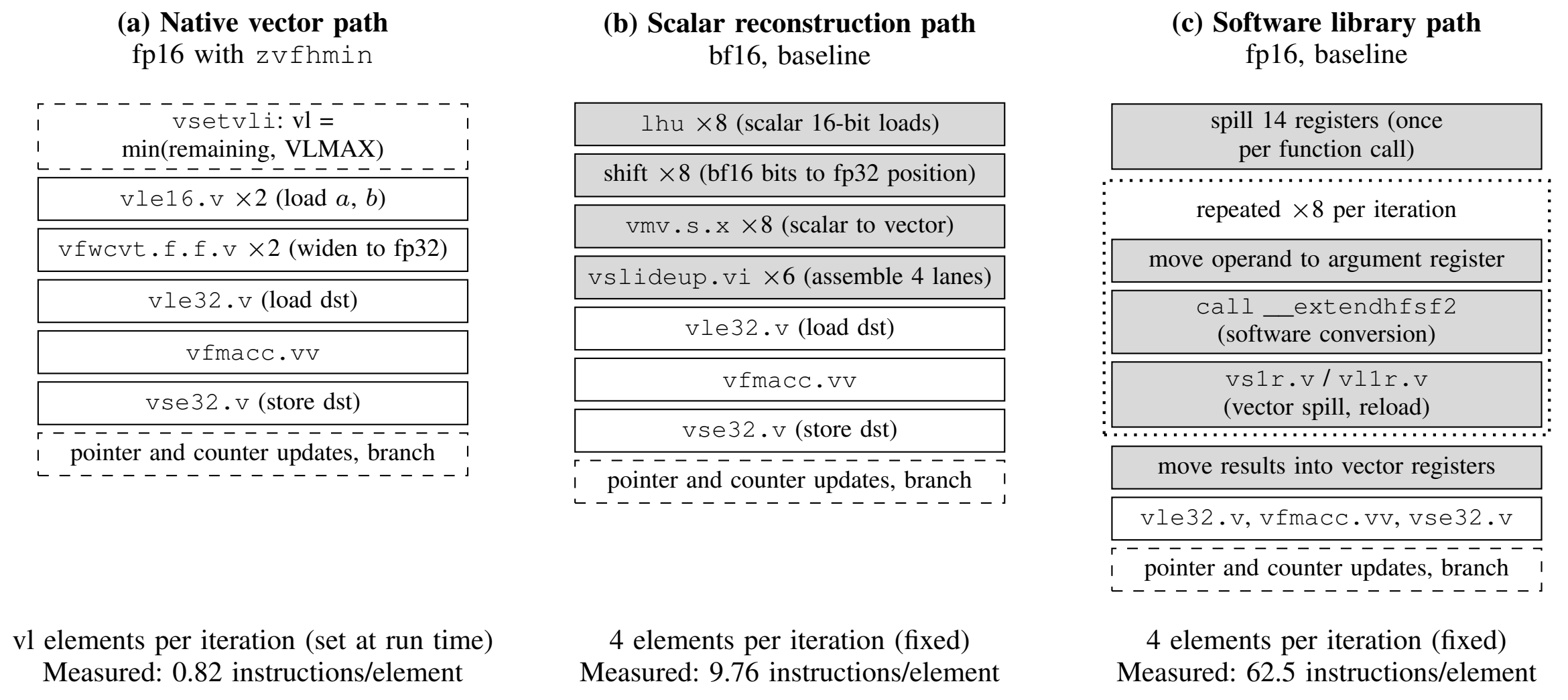
	
	To support the ``degraded'' classification beyond the compiler's 
	own remark text, the instructions in each kernel's loop body are 
	counted and classified as either \emph{productive} (vector loads/stores 
	and the actual arithmetic: \texttt{vle*.v}, \texttt{vse*.v}, 
	\texttt{vfmacc.vv}, \texttt{vwmacc.vv}, \texttt{vfwmaccbf16.vv}, 
	\texttt{vfwcvt.f.f.v}) or \emph{overhead} (everything required only to 
	work around a missing capability: scalar per-element loads, 
	\texttt{vslideup.vi} sequences used to hand-assemble a vector, 
	register spills, or software conversion calls). 
	Fig.~\ref{fig:exec-paths} places the three resulting code paths side 
	by side.
	
	\textbf{bf16, baseline.} The loop body processes a fixed 4 elements 
	per iteration using 39 instructions. Of these, only 3 
	are productive (\texttt{vle32.v}, \texttt{vfmacc.vv}, 
	\texttt{vse32.v}); the rest are 8 scalar half-word loads 
	(\texttt{lhu}), 8 scalar-to-vector element inserts (\texttt{vmv.s.x}), 
	8 shift instructions preparing each half-word, and 6 
	\texttt{vslideup.vi} instructions manually assembling the four 
	scalar-loaded elements into vector registers. The compiler must 
	repeat this overhead on every iteration.
	
	\textbf{bf16, with \texttt{zvfbfmin+zvfbfwma}.} The loop body shrinks 
	to 14 instructions, of which 5 are productive 
	(\texttt{vle32.v}, two \texttt{vle16.v}, \texttt{vfwmaccbf16.vv}, 
	\texttt{vse32.v}); the remaining instructions are address arithmetic 
	and loop control, not per-element reconstruction. The loop also moves 
	from a fixed width of 4 to \texttt{vscale x 4}, so each iteration 
	also covers a number of elements set at runtime (usually larger). 
	Both the per-iteration instruction count and the 
	number of iterations improve at once.
	
	\textbf{fp16, baseline.} This case is more severe. Before the loop 
	begins, the function saves 13 integer registers and one 
	floating-point register to the stack, a sign of high register 
	pressure. The loop body is 114 instructions long and performs 8 explicit calls 
	to the software routine \texttt{\_\_extendhfsf2} (one per scalar 
	element requiring conversion: 4 elements $\times$ 2 operands), each 
	surrounded by register moves and vector-register spill/reload pairs 
	(\texttt{vs1r.v}/\texttt{vl1r.v}) to preserve state across the call. 
	Function-call overhead of this kind (branch-and-link, callee 
	save/restore, return) is much more expensive per element than 
	the register-only shuffling seen in the bf16 baseline.
	
	\textbf{fp16, with \texttt{zvfhmin}.} The loop body drops to 
	17 instructions, replacing all 8 software conversion 
	calls with 2 hardware widening-convert instructions 
	(\texttt{vfwcvt.f.f.v}), one per input array, with no calls, no 
	spills, and no manual element assembly.
	
	\textbf{int8, baseline (\texttt{dot\_int8}).} By comparison, this loop 
	body uses 12 instructions to process an already-scalable 
	(\texttt{vscale x 4}) chunk per iteration, of which 5 are productive 
	(two \texttt{vle8.v} loads, two \texttt{vsext.vf2} widening 
	sign-extends, and \texttt{vwmacc.vv}), with a single 
	\texttt{vredsum.vs} reduction after the loop and no scalar 
	reconstruction of any kind, consistent with its clean vectorization 
	remark and requiring no additional extension flags.
	
	\textbf{int8, baseline (\texttt{fma\_int8}, elementwise).} This 
	elementwise counterpart to \texttt{dot\_int8} uses 15 instructions 
	to process a \texttt{vscale x 4} chunk per iteration, of which 7 are 
	productive: two \texttt{vle8.v} loads, two \texttt{vsext.vf2} 
	widening sign-extends, one \texttt{vle32.v} load of the accumulator, 
	one \texttt{vwmacc.vv} widening multiply-accumulate, and one 
	\texttt{vse32.v} store. The rest are address arithmetic and loop 
	control, with no scalar reconstruction. This matches the clean 
	outcome observed for \texttt{dot\_int8} despite the different loop 
	structure (elementwise update versus reduction), confirming that int8's 
	clean vectorization is not a side effect of the reduction structure but 
	holds for the same elementwise shape used by \texttt{fma\_f32}, 
	\texttt{fma\_bf16}, and \texttt{fma\_fp16}.
	
	\textbf{fp32, baseline (reference).} This loop body uses 13 
	instructions to process a \texttt{vscale x 4} chunk per iteration, of 
	which 5 are productive: three \texttt{vle32.v} loads (both inputs and 
	the accumulator), one \texttt{vfmacc.vv}, and one \texttt{vse32.v} 
	store. The rest are address arithmetic and loop control, with no 
	scalar reconstruction of any kind. This 
	serves as the reference case against which the narrow-type results 
	are compared: \texttt{fma\_bf16} and \texttt{fma\_fp16}, under baseline 
	flags, do not reach this level of code quality despite processing 
	narrower data, while \texttt{dot\_int8}, \texttt{fma\_int8}, and the 
	extension-enabled \texttt{fma\_bf16}/\texttt{fma\_fp16} do.
	
	These counts cover each vector loop body, from its label to its 
	closing branch, in the static assembly. They exclude the scalar loop 
	that handles leftover elements in the fixed-width cases. Their purpose is to make the severity 
	ordering concrete and checkable against the 
	assembly: \texttt{fp32} $\approx$ \texttt{int8} (both \texttt{dot\_int8} 
	and \texttt{fma\_int8}) $<$ \texttt{bf16 (baseline)} $<$ 
	\texttt{fp16 (baseline)}, with both floating-point narrow cases 
	resolved by the corresponding extension. The next subsection measures 
	the same ordering dynamically.
	
	\subsection{Measured execution cost}
	
	\begin{table}[t]
		\centering
		\caption{Measured execution under QEMU (median of 5 runs, 14.7M elements per run)}
		\label{tab:exec}
		\setlength{\tabcolsep}{3.5pt}
		\begin{tabular}{lrrrr}
			\toprule
			\textbf{Binary} & \textbf{Insns (M)} & \textbf{Insns/elem} & \textbf{vs.\ fp32} & \textbf{Wall (s)} \\
			\midrule
			\texttt{fma\_f32} & 8.60 & 0.58 & 1.00$\times$ & 0.62 \\
			\texttt{fma\_int8} & 10.24 & 0.69 & 1.19$\times$ & 0.60 \\
			\texttt{dot\_int8} & 8.40 & 0.57 & 0.98$\times$ & 0.28 \\
			\texttt{fma\_bf16} base & 143.94 & 9.76 & 16.74$\times$ & 1.44 \\
			\texttt{fma\_bf16} ext & 8.47 & 0.57 & 0.98$\times$ & 1.01 \\
			\texttt{fma\_fp16} base & 921.87 & 62.52 & 107.19$\times$ & 15.37 \\
			\texttt{fma\_fp16} ext & 12.15 & 0.82 & 1.41$\times$ & 0.95 \\
			\bottomrule
		\end{tabular}
	\end{table}
	
	\begin{figure}[t]
		\centering
		\begin{tikzpicture}
	\begin{axis}[
		ybar, ymode=log, log origin=infty,
		width=\columnwidth, height=5.4cm,
		bar width=10pt, bar shift=0pt,
		ymin=0.2, ymax=300,
		ylabel={Dynamic instructions per element},
		ylabel style={font=\footnotesize},
		symbolic x coords={fp32,int8,dotint8,bf16b,bf16e,fp16b,fp16e},
		xtick={fp32,int8,dotint8,bf16b,bf16e,fp16b,fp16e},
		area legend,
		xticklabels={\texttt{fp32}, \texttt{int8}, \texttt{dot\_int8}, \texttt{bf16} base, \texttt{bf16} ext, \texttt{fp16} base, \texttt{fp16} ext},
		x tick label style={rotate=35, anchor=east, font=\scriptsize},
		y tick label style={font=\scriptsize},
		ymajorgrids, grid style={gray!30},
		enlarge x limits=0.08,
		nodes near coords, point meta=explicit symbolic,
		nodes near coords style={font=\scriptsize},
		legend style={font=\scriptsize, at={(0.02,0.98)}, anchor=north west},
		]
		\addplot[draw=black, fill=white] coordinates {
			(fp32,0.583) [0.58] (int8,0.695) [0.69] (dotint8,0.570) [0.57]
			(bf16e,0.574) [0.57] (fp16e,0.824) [0.82]};
		\addplot[draw=black, fill=gray!55] coordinates {
			(bf16b,9.762) [9.76] (fp16b,62.52) [62.5]};
		\legend{Clean vectorization, Degraded (baseline)}
	\end{axis}
\end{tikzpicture}
		\caption{Dynamic instructions executed per array element (log scale), from Table~\ref{tab:exec}.}
		\label{fig:insn-per-elem}
	\end{figure}
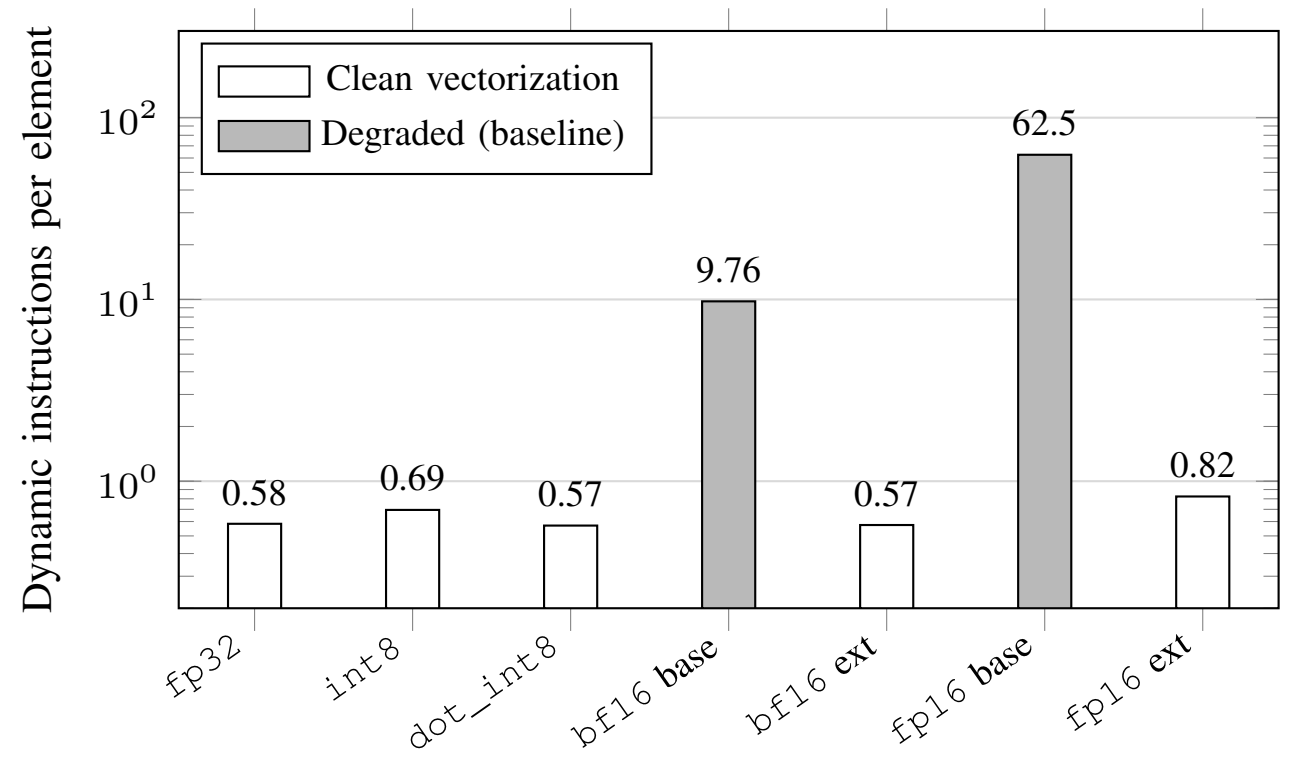
	
	Table~\ref{tab:exec} and Fig.~\ref{fig:insn-per-elem} report the 
	measured results. Run-to-run variation was below 0.001\% in 
	instruction count and below 9\% in wall-clock time. The baseline 
	\texttt{fma\_bf16} executes 17.0$\times$ as many instructions as its 
	extension-enabled build, and the baseline \texttt{fma\_fp16} executes 
	75.9$\times$ as many as its extension-enabled build. With the 
	extensions, both narrow floating-point kernels execute within 
	1.41$\times$ of the fp32 kernel's instruction count, and the bf16 
	build executes slightly fewer instructions than fp32, while reading 
	half as many input bytes. Without them, bf16 executes 16.7$\times$ and 
	fp16 107$\times$ the instructions of fp32 for the same elements.
	
	The measured counts are consistent with the static counts. The 
	baseline bf16 loop body of 39 instructions per 4 elements 
	corresponds to 9.75 instructions per element, against a 
	measured 9.76. The baseline fp16 loop body of 114 instructions per 
	4 elements accounts for 28.5 of the measured 62.5 instructions per 
	element; the rest comes from the two calls to 
	\texttt{\_\_extendhfsf2} per element, about 17 instructions each. The clean kernels execute 
	between 0.57 and 0.82 instructions per element, because each 
	scalable-width iteration covers many elements at VLEN = 256.
	
	Host wall-clock time follows the same ordering for the baseline 
	builds: 1.43$\times$ for bf16 and 16.2$\times$ for fp16, baseline over 
	extension. It is not a reliable measure of relative cost, however. The 
	bf16 extension build executes 1.6\% fewer instructions than fp32 but 
	takes 1.63$\times$ as long on the host, meaning QEMU's implementation 
	of \texttt{vfwmaccbf16.vv} costs more host time per instruction than 
	its fp32 counterpart. This reflects the emulator's implementation, not 
	RISC-V hardware.
	
	\section{Discussion}
	
	The results show that narrow AI-relevant types are not vectorized 
	equally well under a single \texttt{-march=rv64gcv} target. Integer narrow types (\texttt{int8} 
	widening to \texttt{int32}) vectorize as cleanly as standard-width \texttt{fp32}, with no 
	additional flags, because integer widening is covered by the base 
	vector extension. Floating-point narrow types (\texttt{bf16}, 
	\texttt{fp16}) instead require separate, easily 
	overlooked opt-in extensions (\texttt{zvfbfmin+zvfbfwma}, 
	\texttt{zvfhmin}) to reach the same quality of vectorization; without 
	them, the compiler still emits a vector loop, but pads it with scalar 
	work to reconstruct or convert operands one element at a time. In 
	the fp16 case, it falls back to a software conversion routine per 
	element, a much more expensive failure than bf16's 
	in-register shuffling.
	
	For developers deploying AI models, the practical consequence is that storing weights 
	in bf16 or fp16 to save memory can make the hot loop slower than 
	leaving them in fp32. Under the baseline target, the bf16 kernel 
	executes 16.7$\times$ the instructions of the fp32 kernel while reading 
	half the input bytes, and nothing in the build output, the program's 
	results, or its default diagnostics shows this. With the 
	extensions enabled, the narrow types deliver the expected benefit: 
	fp32-level instruction counts with half the memory footprint.
	
	This difference is not made clear to a 
	developer relying on compiler diagnostics. The vectorizer correctly 
	detects and reports degraded vectorization for both narrow 
	floating-point types, but does so using remark text that does not 
	distinguish a \emph{fixable} limitation (missing extension support, 
	resolved entirely by adding a flag) from an \emph{inherent} one 
	(the unavoidable cost of a precision-changing cast). Both are reported 
	through the same \texttt{-Rpass-analysis=loop-vectorize} channel with 
	similar phrasing. A developer encountering the ``not supported for all 
	element types'' remark has no textual indication that a two-flag change 
	would resolve it, and could reasonably conclude that degraded 
	performance is an inherent property of the workload rather than a 
	missing target-feature flag. As Fig.~\ref{fig:llvm-pipeline} shows, the 
	remark originates from a single legality check in 
	\texttt{RISCVTargetTransformInfo} that depends only on the enabled 
	extensions, so the compiler has the information needed to name the 
	missing extension in the remark. Clang already attaches fix-it hints 
	to front-end errors when the correction is clear; a remark naming 
	the missing extension would apply the same idea to optimization 
	remarks, which matters because messages that are harder to read lead 
	developers to worse results \cite{barik2017}. This gap has been 
	reported to the LLVM project, along with a proposal to name the 
	rejected type in the remark \cite{llvmissue}.
	
	The overhead observed in the unpatched bf16 and fp16 cases consists 
	partly of independent scalar operations, which a wide enough 
	out-of-order core could in principle run in parallel, partly 
	offsetting the instruction-count penalty. However, this offset is 
	incomplete for three reasons: instruction fetch/decode bandwidth is 
	used up regardless of execution-level parallelism; the overhead does 
	not scale with vector width the way productive vector instructions do, 
	so the relative gap widens on hardware with wider vector units; and 
	part of the overhead (the sequential \texttt{vslideup.vi} chain in the 
	bf16 case, and the call overhead in the fp16 case) is inherently 
	serial and cannot be parallelized regardless of core width. This 
	offset is also weaker in practice for the edge-inference setting 
	that motivates this study, where RISC-V implementations are often 
	simpler, narrower-issue cores with less instruction-level 
	parallelism than desktop-class hardware.
	
	\section{Developer Tips}
	
	The following checks apply to developers building AI inference code 
	for RVV hardware with Clang.
	
	\textbf{Compiler version.} Scalable vectorization of 
	\texttt{half} and \texttt{bfloat} under \texttt{Zvfhmin} and 
	\texttt{Zvfbfmin} alone was added to LLVM in November 2024 
	\cite{llvmpr115272}. With older releases, adding these extensions to 
	\texttt{-march} may not produce the scalable loops measured here, so 
	a compiler release that includes this change is needed.
	
	\textbf{Matching \texttt{-march} to the hardware.} On Linux, the 
	\texttt{isa} line of \texttt{/proc/cpuinfo} lists the extensions the 
	processor implements. If it includes \texttt{zvfhmin}, 
	\texttt{zvfbfmin}, or \texttt{zvfbfwma}, these should be added to the 
	\texttt{-march} string. They should not be added otherwise, since a 
	binary that uses these instructions stops with an illegal-instruction 
	fault on hardware that lacks them. For binaries distributed to 
	hardware that is not known in advance, Clang supports function 
	multi-versioning on RISC-V through the \texttt{target\_clones} 
	attribute, which builds several versions of a function and picks 
	one at run time.
	
	\textbf{Profile coverage of bf16.} Targeting the RVA23U64 
	profile enables \texttt{Zvfhmin}, which resolves the fp16 case, but 
	\texttt{Zvfbfmin} and \texttt{Zvfbfwma} are optional in that profile 
	\cite{rva23} and must be requested separately. Hardware support is 
	also limited: llama.cpp's \texttt{Zvfbfwma} kernels have so far been 
	tested only under QEMU \cite{ggml18199}. On hardware without these 
	extensions, bf16 loops take the baseline path measured in this paper.
	
	\textbf{Remarks for performance-critical kernels.} Building 
	performance-critical files with \texttt{-Rpass=loop-vectorize 
	-Rpass-analysis=loop-vectorize} shows the degradation without 
	reading assembly. Two outputs indicate a missing 
	extension: the remark ``Scalable vectorization is not supported for 
	all element types found in this loop,'' and a fixed vectorization 
	width such as \texttt{4} where \texttt{vscale x N} is expected. The 
	remark ``floating point conversion changes vector width'' appears with 
	or without the extension and does not by itself indicate a missing 
	flag.
	
	\textbf{Conversion calls in the assembly.} Compiling a kernel 
	with \texttt{-S} and searching for \texttt{call \_\_extendhfsf2} (or 
	\texttt{\_\_truncsfhf2} for the reverse conversion) identifies fp16 
	loops that convert in software. For bf16, the signature is a loop with 
	per-element \texttt{lhu} loads and \texttt{vslideup.vi} sequences and 
	no \texttt{vfwmaccbf16.vv} or \texttt{vfwcvtbf16.f.f.v}.
	
	\textbf{Storage type and target.} If the 
	target lacks the extensions, the measured options for this loop are: 
	fp32 at 0.58 instructions per element with twice the memory of bf16; 
	int8 at 0.57 to 0.69 instructions per element, since integer widening 
	needs only the base vector extension; bf16 at 9.76; and fp16 at 62.5. 
	Converting bf16 or fp16 weights to fp32 once at load time moves the 
	conversion out of the inner loop at the cost of memory.
	
	\textbf{Measurement under QEMU.} Wall-clock 
	time under QEMU reflects the emulator's implementation of each 
	instruction, and in this study it ranked the bf16 extension build 
	behind fp32 despite nearly equal instruction counts. QEMU's 
	instruction-count plugin gives a more stable comparison of code paths.
	
	\section{Limitations}
	
	This study has several limitations. First, it evaluates a single 
	compiler, Clang. The static results were checked with three LLVM 
	versions, but the measured binaries were built only with 
	Clang~22.1.8. Second, execution was measured under emulation, not on RISC-V 
	hardware. Dynamic instruction counts weight all instructions equally 
	and do not model pipelines, caches, memory bandwidth, or the per-element 
	cost of vector instructions on a particular implementation, and 
	results were collected at a single vector length (VLEN = 256). Third, 
	the absolute count for the baseline fp16 binary depends on the 
	implementation of \texttt{\_\_extendhfsf2} linked into it; a different 
	runtime library would change the count, though not the number of calls 
	per element. Fourth, the kernel set is narrow (one or two 
	FMA-shaped kernels per type) and does not establish how broadly this 
	pattern holds across other loop shapes (reductions beyond \texttt{dot\_int8}, 
	strided access, larger AI kernels such as full GEMM tiles). The 
	question of whether the same asymmetry holds on GCC's RISC-V 
	vectorizer is also not addressed, so it is not known whether this is an 
	LLVM-specific gap or a broader one across compilers.
	
	\section{Conclusion}
	
	This study finds that LLVM's RISC-V auto-vectorizer does not handle 
	AI-relevant narrow numeric types equally well under a single 
	\texttt{-march=rv64gcv} target: int8 kernels vectorize cleanly by 
	default, while bf16 and fp16 fused multiply-accumulate kernels degrade 
	to increasingly scalar-heavy fallback code unless additional 
	target-feature extensions are enabled. Under emulation, the 
	degraded bf16 and fp16 kernels execute 17.0$\times$ and 75.9$\times$ 
	the instructions of their extension-enabled builds, and the 
	extension-enabled builds come within 1.41$\times$ of fp32. The 
	fallback is triggered by the element-type legality check the 
	vectorizer performs through \texttt{RISCVTargetTransformInfo}. Compiler diagnostics 
	accurately report degradation but do not distinguish between a limitation that a compiler flag fixes and 
	one that reflects an inherent cost of the operation, giving developers no direct signal toward the fix. 
	
	Future work could extend this analysis by 
	measuring the same kernels on physical RISC-V hardware that implements 
	\texttt{Zvfhmin} and \texttt{Zvfbfwma}, and at more than one vector 
	length; evaluating additional loop shapes beyond fused 
	multiply-accumulate, such as reductions or strided access patterns 
	common in convolution and attention kernels; and testing whether the 
	same diagnostic ambiguity is present in GCC's RISC-V vectorizer, to 
	determine whether this is an LLVM-specific gap or a broader pattern 
	across RISC-V compiler toolchains.

\end{document}